\documentclass[%
reprint,amsmath,amssymb,aps,nofootinbib]{revtex4-1}

\usepackage{amsmath}
\usepackage{amssymb}
\usepackage[normalem]{ulem}
 \usepackage{graphicx}
 \usepackage{subcaption}
 \usepackage{hyperref}
\graphicspath{ {./} }  

\newcommand{\beq}{\begin{equation}}
\newcommand{\eeq}{\end{equation}}
\def\barr{\begin{array}}
\def\earr{\end{array}}

\newcommand{\bsp}{\begin{split}}
\newcommand{\esp}{\end{split}}
\newcommand{\bit}{\begin{itemize}}
\newcommand{\eit}{\end{itemize}}
\usepackage{color}
\definecolor{darkcyan}{cmyk}{1,0,0,0.4}

\begin{document}

%\begin{frontmatter}

\title{Clockwork Neutrinogenesis: Baryogenesis from theory space}

\author{Suvam Maharana}
\email{smaharana@physics.du.ac.in}
\author{Tripurari Srivastava}
\email{tripurarisri022@gmail.com}
\affiliation{Department of Physics and Astrophysics, University of Delhi, Delhi, 110007, India}

%\date{\today}

\begin{abstract}
We propose a minimal clockwork model to illustrate the possibility of baryogenesis via leptogenesis in a theory space setting. The standard lepton sector is augmented with three copies of a clockwork lattice made of SM neutral fermions. The two boundaries of these one-dimensional lattices are endowed with couplings to the SM leptons and three dark sector fermions, respectively. Small neutrino masses and a resonance enhanced Dirac leptogenesis are naturally obtained for anarchic textures of the Yukawa matrices, with $\sim \mathcal{O}(1)$ elements, provided the heavy clockwork fermions have masses $\gtrsim \mathcal{O}(10 \, \mbox{TeV})$.
\end{abstract}

\maketitle

%\end{frontmatter}
\section{Introduction}
The origins of the apparent lightness of the neutrinos as well as the observed baryon asymmetry of the Universe (BAU), remain elusive. This has elicited a myriad of investigations over the years towards combined explanations of neutrino masses and the BAU, most popularly through the mechanism of baryogenesis via leptogenesis~\cite{Fukugita:1986hr,Pilaftsis:2003gt,Buchmuller:2004tu,Buchmuller:2005eh}. These scenarios commonly employ the seesaw mechanism \cite{Davidson:2008bu} wherein the Standard Model (SM) Lagrangian is augmented by two or more heavy neutral fermions (\emph{e.g.} in type-I seesaw) and the lightness of the active neutrinos is obtained through a suppression generated by the mass scale of the new fermions, typically via relations like
\beq \nonumber
m_{\nu} \sim y^2 \frac{v_{EW}^2}{M_N} \, ,
\eeq
where $y$ denotes a dimensionless Yukawa coupling, $v_{EW}$ characterizes the electroweak scale and $M_N$ specifies the mass scale of the heavy neutrinos. This suggests that in order to obtain acceptable neutrino masses ($m_{\nu} \lesssim 0.1$ eV) with $\mathcal{O}(1)$ Yukawa couplings one requires atleast one of the heavy neutrinos to have a characteristic mass scale hierarchically larger than the EW scale. Of course, with $y \ll \mathcal{O}(1)$ the scale $M_N$ can be brought closer to the EW scale. Although this is naturally realised in radiative neutrino mass models \cite{Cai:2017jrq,Gu:2007ug} where the effective operator responsible for mass generation is radiatively induced\footnote{An interesting alternative can also be found in braneworld scenarios where hierarchical masses and lepton asymmetry are generated through a geometry-induced exponential warping --- e.g., having neutrinos in the bulk of a 5D Randall-Sundrum geometry~\cite{Medina:2006hi,Gherghetta:2007au}.}, a low-scale leptogenesis is often found difficult to achieve in minimal scenarios (\emph{e.g.} in the Ma model \cite{Ma:2006km}) unless small Yukawa couplings are assumed.

In this work, we investigate the possibility of a unified explanation for the neutrino masses and leptogenesis (\emph{neutrinogenesis}) without invoking small Yukawa couplings or large hierarchies in the new mass scales. For that, we veer off the usual route and approach the problem from a theory space perspective. To this end, the \textit{clockwork} (CW) paradigm \cite{Giudice:2016yja,Kaplan:2015fuy,Choi:2015fiu} provides an attractive framework to naturally generate hierarchical couplings in a theory with $\mathcal{O}(1)$ parameters through localization effects in the theory space.
Neutrino mass models based on the clockwork mechanism have been studied before in refs.\cite{Giudice:2016yja,Hambye:2016qkf,Ibarra:2017tju,Hong:2019bki,Kitabayashi:2019qvi}. These models typically invoke a chain of Weyl fermions, all neutral under the SM gauge group, which define a lattice in the theory space. Each clockwork fermion mixes with its nearest neighbour(s) with a weight $q$. The SM neutrinos then couple to one of the right-handed lattice fermions, or, in the language of theory spaces, are localised at one of the \textit{sites} (minimally via the SM Higgs). The mechanism then dictates that the effective Yukawa couplings are suppressed by a factor $q^{-N}$, where $N$ denotes the number of sites. The mass spectrum of such models consist of a band of heavy neutral fermions, closely separated in mass,  in addition to the light neutrinos. Evidently, the mass scale of the heavy neutrinos in this setup is not tied to the neutrino mass generation, unlike the case in seesaw models, which enables the prospect of realizing a low-scale leptogenesis with $\mathcal{O}(1)$ couplings and no large hierarchies among the heavy neutrino masses. The possibility of leptogenesis within a clockwork construction has been discussed before, although not explicitly investigated, in ref.\cite{Hambye:2016qkf} in the context of a standard lepton number violating setup.

  We demonstrate, explicitly,  the possibility of a low-scale Dirac (lepton number conserving) leptogenesis ~\cite{Dick:1999je,Murayama:2002je,Cerdeno:2006ha} within a minimal CW setup for generating small neutrino masses. The choice of a lepton number conserving scenario is motivated by the fact that the fermion mass terms in a CW theory are characteristically Dirac in nature, and also by the absence of any telltale evidence indicating a Majorana nature of the neutrinos. The neutrinogenesis in this case is achieved with a completely \emph{anarchic} Higgs Yukawa matrix in the neutrino sector comprising of nearly $\mathcal{O}(1)$ entries, where we use the term ``anarchic" in the manner of ref.\cite{Hall:1999sn}, signifying Yukawa matrices having comparable entries, exhibiting no discernible patterns or hierarchies.
 This requires introduction of three fermionic clockwork chains, one for each flavor of the SM neutrinos, and three flavors of a new light neutral fermion to ensure an observable CP violation. To avoid unnecessary inter-flavor hierarchies in the heavy CW spectrum, a simple $\mathbb{Z}_3$ exchange symmetry is imposed among the three flavors which is broken solely by the Yukawa couplings of the light fermions (SM leptons and the extra fermions) to the clockwork sector. The required CP asymmetry for leptogenesis is then obtained by virtue of a resonance enhancement effected by the resulting quasi-degenerate masses of the heavy neutral fermions. The required mass-splittings for the resonance condition are obtained quite naturally, independent of the textures of the Yukawa matrices, via radiative corrections to the heavy fermion mass terms originating from the explicit breaking of the flavor symmetry. However, this is true only when the out-of-equilibrium dynamics of leptogenesis concludes before electroweak symmetry breaking (EWSB) occurs. With this constraint, we find that the heavy neutrinos must have masses $\gtrsim \mathcal{O}(10 \,  \mbox{TeV})$.
\section{Model}
\subsection{Clockwork with fermions}
We define the CW theory space with $N$ copies of left and $N+1$ copies of right-handed Weyl fermions --- all singlets under the SM gauge group --- and the Lagrangian \cite{Giudice:2016yja},
\beq 
\begin{split}
\mathcal{L}_{CW} = \mathcal{L}_{kin}- m\sum_{j=0}^{N-1}\bar{\psi}_{L j}\left(\psi_{R_{j}}-q \psi_{R_{j+1}}\right) + \mbox{h.c.} \, .
\end{split}
\label{eq:lagcw}
\eeq
Here, $m$ is a mass parameter and $q$ is a dimensionless parameter which enables the CW mechanism for values $q > 1$ \footnote{Note that $m$ is chosen to be real without the loss of generality and so is the parameter $q$ for brevity.}. The combination $mq$, therefore,  approximates the characteristic scale of the physical spectrum for large $q$. The near-neighbor mixing terms break the full chiral symmetry $U(N)_L \times U(N+1)_R$ of the kinetic terms into a residual factor $U(1)_R$, meaning that the physical spectrum contains one massless right-handed state. %[What about minor fluctuations in the parameters ? Possible UV completion ?] 
The massive eigenvalues of the spectrum, on the other hand, are given by,
\beq
m_{n>0}= m \sqrt{1+q^{2}-2q\cos{[n\pi/(N+1)]}} \sim mq .
\label{eqn:evscal}
\eeq
With the unphysical $(\psi)$ $\leftrightarrow$ mass basis $(\Psi)$ transformation defined as $\Psi_n = \sum_{j} a_{n j} \psi_j$, the rotation matrices (for large $N$) are given by, 

\beq \label{eq:eigv}
\begin{split}
& a_{0j}^R \sim q^{-j}, \, a_{nj}^{L} \sim \sqrt{2/N}\sin{[((j+1) n \pi) /(N+1)]}, \\
& a_{nj}^{R} \sim \sqrt{\frac{2 m^2}{N m^2_n}}\left( q \sin{\frac{j n \pi}{N+1}} - \sin{\frac{(j+1)n \pi}{N+1}}\right).
\end{split}
\eeq
This clearly shows that the right-handed massless state $\Psi_{R 0}$ is localised towards the $j=0$ site, thereby illustrating the clockwork mechanism. The massive modes, on the other hand, are delocalised over the lattice with roughly $\mathcal{O}(1)$ weights. 

\subsection{A model for neutrinos and leptogenesis}

For a minimal realisation of neutrino masses and leptogenesis, we introduce three copies of the aforementioned SM singlet clockwork chains --- one for each flavor of the SM neutrinos --- and three flavors of additional neutral fermions $\chi_{L,R}$ along with a SM singlet real scalar $\Phi$. We then posit the following lepton number conserving Lagrangian which specifies the clockwork sector's interaction with both the SM lepton doublets $L$ and the $\chi$'s through \emph{localised} Yukawa couplings,
\beq \label{eq:lagcwsm}
\begin{split} 
&\mathcal{L}_{model} = \mathcal{L}_{SM} +\sum_{\beta=1}^{3} \mathcal{L}^{\beta}_{CW} + \mathcal{L}_{CW-SM} + \mathcal{L}_{CW-DS} + \mbox{h.c.} \, ,\\
&\mathcal{L}_{CW-SM} \equiv  - \sum_{\alpha, \beta}\lambda_{1 \alpha \beta} \Bar{L}_{\alpha}\tilde{H}\psi_{R N}^{\beta} \, ,\\
&\mathcal{L}_{CW-DS} \equiv - \sum_{\alpha, \beta} \lambda_{2 \alpha \beta} \Phi \, \Bar{\psi}_{L 0}^{\beta} \, \chi_{R \alpha} - \sum_{\alpha, \beta} \lambda^{\chi}_{ \alpha \beta} \Phi \, \Bar{\chi}_{L \beta} \, \chi_{R \alpha} \, ,
\end{split}
\eeq
where $\lambda^{1,2,\chi}_{\alpha \beta}$ are the Yukawa couplings corresponding to the flavor indices $\alpha, \beta \in [1,3]$. Here, $\tilde{H}$ has the usual definition $\tilde{H}=i \sigma_2 H^*$. Therefore, the model is described by two components, $\mathcal{L}_{CW-SM}$ and $\mathcal{L}_{CW-DS}$, beyond the SM Lagrangian $\mathcal{L}_{SM}$ which specify the interaction of the CW sector with the SM leptons and the $\chi-\Phi$ sector (which constitutes a dark sector as justified later), respectively. While $\mathcal{L}_{CW-SM}$ is sufficient to generate the light neutrino masses within such a lepton number conserving setup, the Yukawa matrix $\lambda_1$ (with complex entries) alone fails to account for a nonzero CP asymmetry as required for leptogenesis through the decay of the heavy CW fermions $\Psi_n$ (to be discussed shortly). To remedy this, what is required is the presence of another Yukawa matrix which does not commute with $\lambda_1$. The component $\mathcal{L}_{CW-DS}$, comprising of additional Yukawa interactions enabled by the fields $\chi$ and $\Phi$, serves just this purpose. Note that the assumed field and coupling configuration of the dark sector is not unique to the CW setup and one can, in principle, invoke other configurations as well. However, the simple choice made in the model defined in eq.\ref{eq:lagcwsm} suffices to illustrate the salient features of Dirac leptogenesis within the clockwork paradigm.
\begin{figure}[htb] 
\centering
      \includegraphics[width=0.46\textwidth,height=0.16\textwidth]{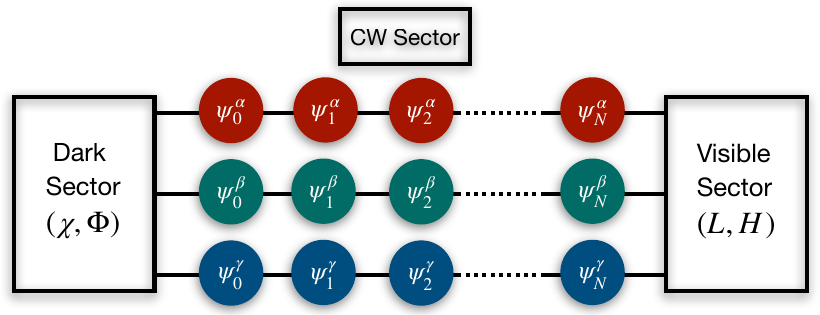}
	\caption{\textit{A schematic depiction of the model.}} \label{fig:schematic}  
\end{figure}

An exchange symmetry is assumed among the three flavors of the CW fermions in $\mathcal{L}_{CW}$, namely, a symmetry under the $\mathbb{Z}_3$ transformations $\psi_{\alpha} \leftrightarrow \psi_{\beta}$, only to be broken by the flavor mixing Yukawa couplings $\lambda_{1,2}$. This simply means that the set of parameters $(m,\, q,\, N)$ assume the same values for all the flavors. It will be shown shortly that the assumed configuration of the $\mathbb{Z}_3$ symmetry breaking leads to a loop-induced mixing between the CW flavors which ensures that they are quasi-degenerate with just the right magnitude of mass-splitting so as to realise resonant leptogenesis. The SM leptons interact with the CW sector through Yukawa couplings ($\lambda_1$) with the fermions $\psi_{R N}^{\beta}$. With reference to eq.\ref{eq:eigv}, this localised coupling will eventually lead to small masses for the SM neutrinos through the exponential suppression effected by the distribution of the massless right-handed CW fermion ($\Psi_{R 0}$).   For simplicity and, more importantly, to restrict equilibration of the left- and right-handed CP asymmetries in the leptogenesis mechanism (to be discussed in the ensuing section), we demand that the dynamics of the SM neutrino sector is not significantly affected by terms other than the $\lambda_1$ interaction which is localised at the $j=N$ site. To that end, we connect the dark sector fermions $\chi_{\alpha}$ to the CW sector through a coupling $\lambda_2$ with the field $\psi_{L 0}$ only, \emph{i.e.} at the CW site farthest from where the SM leptons interact with the CW fermions. Furthermore, we restrict terms like $\Phi \psi_{R 0} \chi_L$ by introducing a $\mathbb{Z}_2$ symmetry under which only the fields $\chi_R$ and $\Phi$ have an odd parity \footnote{Note that the $\mathbb{Z}_2$ symmetry still allows the bare mass term $m_{\psi \chi}\psi_{R 0} \chi_L$. However, we assume a mild hierarchy $[m_{\psi \chi}/(\lambda_{\chi}\langle \Phi\rangle)] \lesssim 0.1$, in which case there is no substantial effect on the masses of the light neutrinos as well as the $\chi$'s as obtained from the $\lambda_1$ and $\lambda_{\chi}$ couplings, respectively. Since the term as such would not have any bearing on the leptogenesis mechanism, we have ignored it altogether for simplicity.}. This also prohibits $\Phi$ from interacting with other fields in the CW sector. The $\lambda^{\chi}$ couplings, on the other hand, largely concern the dynamics of the $\chi$ sector alone. For one, the scalar potential for $\Phi$ is assumed to be such that it acquires a nonzero vacuum expectation value (VEV) $\langle \Phi \rangle = v_{\Phi}$ lending masses to $\chi$'s --- $m_{\chi} \sim \lambda_{\chi} v_{\Phi}$. For brevity we assume $\lambda^{\chi}$ to be diagonal. Note that while $\chi_R$ might seem to resemble the field $\nu_R$ usually encountered in Dirac leptogenesis models (see e.g. \cite{Cerdeno:2006ha}), it does not play the same role here. Unlike the case in conventional seesaw-based models, the $\chi_R$ fields do not contribute to the light neutrino masses as such and are relevant mainly for the leptogenesis mechanism. Also, from a phenomenological standpoint, the introduction of the left-handed fields $\chi^{(\alpha)}_L$ was necessary solely for the sake of having Dirac mass terms\footnote{Although a Majorana mass term for $\chi$ would have been a simpler choice, we choose to have only Dirac mass terms in the Lagrangian in the spirit of avoiding lepton number violation in the model. Notwithstanding that, introducing Majorana mass terms in the $\chi$ sector would not induce any major qualitative change in the results deduced from our analysis.} for $\chi$'s.  From a theory space perspective, then, the SM fields are localised at the site $j=N$, whereas $\chi$'s and $\Phi$ are localised at $j=0$ (see Fig.\ref{fig:schematic} for an illustration). While this might seem an \emph{ad hoc} construction at this point, it can be justified, along the lines of ref.\cite{Hambye:2016qkf}, by identifying the symmetries $U(1)_{R N}$ and $U(1)_{L 0}$ with the SM and the $\chi$ lepton numbers, respectively, \emph{i.e.} $U(1)_{R N} \equiv U(1)_{L_{SM}}$ and $U(1)_{L 0} \equiv U(1)_{\chi}$. This naturally enables the localised couplings assumed in eq.\ref{eq:lagcwsm} while restricting terms involving $\psi_{R 0}$ and $\chi_L$. As an alternative realisation, the effective theory can also be envisaged as the deconstruction of a 5D theory with bulk fermions in a (warped) linear dilaton geometry with 3-branes at the boundaries. A discussion on this is presented in \ref{sec:app5dcwm}.

Coming back to the flavor mixing Yukawa interactions, we see that the two $3 \times 3$ complex matrices $\lambda_{1,2}$ break the residual chiral symmetry in each clockwork chain. As a result of the $\lambda_1$ couplings, the massless right-handed fermion in each CW chain pairs with a left-handed SM neutrino to get a Dirac mass following the EWSB. On the other hand, the introduction of $\lambda_2$ facilitates CP violation and, hence, leptogenesis, provided it does not commute with $\lambda_1$. It is important to remark here that the explicit $\mathbb{Z}_3$ breaking terms giving rise to flavor mixing in the model are crucial to the leptogenesis mechanism to be discussed in the subsequent sections. We will show, however, that successful leptogenesis is achievable for random textures of the Yukawa matrices $\lambda_{1,2}$ and, hence, the details of the origin of the $\mathbb{Z}_3$ symmetry breaking are not essential to drawing conclusions from our low energy picture. Even so, a possible realisation of the $\mathbb{Z}_3$ breaking can be found in the aforementioned 5D model with brane-localised flavons breaking the flavor symmetry spontaneously (see \ref{sec:app5dcwm}).

We stress here that consistency with the argument that $H$ and $\Phi$ are localised at different CW sites demands that they do not interact with each other\footnote{In general, a mixing between $H$ and $\Phi$ would have different implications on the cosmological fate of the particles $\chi$ and $\Phi$ than what is discussed here. We do not delve into that aspect here as it is not directly germane to matters related to leptogenesis.}. Therefore, a typical form of the scalar potential would be\footnote{Probable issues related to domain walls may be mitigated by introducing soft $\mathbb{Z}_2$ breaking terms in the $\Phi$ potential.},
\beq
V_{H,\Phi} = m^2_{H}H^2 + m^2_{\Phi}\Phi^2 + \lambda_H \left(H^{\dagger}H \right)^2 + \lambda_{\Phi} \Phi^4.
\eeq

Since $\Phi$ and $\chi$'s couple only feebly with the SM, mediated by the heavy CW fermions, the lightest among them would typically be rendered stable over cosmological timescales. With this is mind we discuss in a subsequent section the viability of a dark matter candidate in the model. In the minimal setup, however, we find that only a significantly hot dark matter $(\Phi)$ may exist which would, ostensibly, be in conflict with structure formation in the Universe. Thus, it can only constitute a small fraction of the total thermal relic, which is naturally ensured in our scenario for $m_{\chi} \sim m_{\Phi} < \mathcal{O}(1 \, \mbox{MeV}) $. This point is further elucidated in a later section. As concerns leptogenesis, it suffices to have the parameters $\lambda_{\chi}$, $\lambda_{\Phi}$ and $m_{\Phi}$ suitably fixed such that the \emph{dark sector} particles $(\chi \, \mbox{and} \, \Phi)$ are substantially lighter than the CW neutrinos.

Post EWSB, the active neutrino masses and the Yukawa matrix in the flavour basis are related by the bi-unitary transformation,
\beq \label{eq:lam1}
U_L . m^{\nu}_D . U_R^{\dagger} \sim \left[q^{-N} v_H/\sqrt{2}\right]\lambda_1,
\eeq
where $v_H$ is the Higgs VEV, $U_{L,R}$ are the transformation matrices for the left and right handed fields, respectively, and $m^{\nu}_D$ denotes the diagonal neutrino mass matrix. With the usual assumption that the charged leptons are flavor diagonal, $U_L$ becomes the conventional (Pontecorvo-Maki-Nakagawa-Sakata) PMNS matrix parametrised with three mixing angles and a phase. The matrix $U_R$, on the other hand, has nine independent parameters as dictated by unitarity. The factor $q^{-N}$ corresponds to the overlap of the $n=0$ right-handed CW fermion with the unphysical field $\psi_{R N}$.  A straightforward scan of the CW parameters shows that the Higgs Yukawa $\lambda_1$ can have $\mathcal{O}(1)$ entries while being consistent with the oscillation data~\cite{Esteban:2020cvm} for $q \geq 2$ and $N \sim \mathcal{O}(10)$. However, one ought to examine whether such a scenario is adequate for baryogensis as well. We show in the ensuing sections that this is indeed true.

\section{Clockwork Leptogenesis}
\begin{figure}[htb] 
\centering
\includegraphics[scale=0.7,keepaspectratio=true]{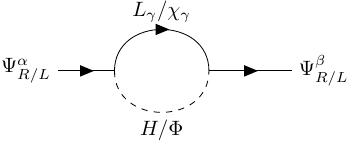}
	\caption{\textit{One-loop interflavor mixing.}} \label{fig:mixing}  
\end{figure}

 With the flavor exchange symmetry being broken by the couplings $\lambda_{1,2}$, the CW fermions of different flavors mix radiatively. The contributing diagrams, before EWSB, are as shown in Fig.\ref{fig:mixing}. The induced mixing at the CW level $(n)$ is characterised by the loop factor
\beq \label{eq:splitloop}
\begin{split}
 \delta^{(\alpha,\beta)}_{n}  \sim \frac{1}{16\pi^2}& \sum_{\gamma}\Big[2\lambda_{1 \alpha \gamma}\lambda^{*}_{1 \gamma \beta}  a^{(\alpha)}_{R \, n N}a^{(\beta)}_{R \, n N} \\
 &+ \lambda_{2 \alpha \gamma}\lambda^{*}_{2 \gamma \beta} a^{(\alpha)}_{L \, n 0}a^{(\beta)}_{L \, n 0}\Big]\ln{\frac{M^2}{(mq)^2}},
\end{split}
\eeq  
where, $a_{L,R}$ are the unperturbed distributions for the CW eigenstates and $M$ is assumed to be a cutoff scale at which the one-loop mixing $\delta$ vanishes \footnote{Naively, this is a particular choice for the initial value condition stipulating the renormalisation scale dependence of $\delta$. The condition, of course, would be sensitive to the details of the possible UV completions.}. In other words, $M$ marks the scale beyond which the exchange $\mathbb{Z}_3$ symmetry is restored. Because of the logarithmic dependence, $M$ can be as large as $\mathcal{O}(100 \, \mbox{TeV})$ without affecting the results qualitatively. This results in a small mass-splitting between the hitherto degenerate fermions. For an anarchic texture of the Yukawa matrices $\lambda_{1,2}$, it is not difficult to infer that the mass-splitting between any two degenerate heavy neutrinos of CW level $(n)$ and belonging to the flavors $(i,j)$ in the mass-basis would be comparable to their decay widths, \emph{i.e.},
\beq
\Delta m^{(i,j)}_n \sim \Gamma\left( \Psi_{i,j}\right).
\label{eq:delmeq}
\eeq
What is interesting here is that the above relation is valid quite generally and does not require any \emph{ad hoc} fine-tuning of the model parameters. Consequently, the resonance enhancement works even for anarchic textures of the Yukawa matrices. The only underlying assumption here is of the flavor symmetry in the CW sector. 
We note that this feature is not endemic to the CW construction but shares its attributes with other resonant leptogenesis scenarios employing radiatively induced mass-splittings (see \emph{e.g.} refs.\cite{Hambye:2004jf,Blanchet:2009bu}).
It is worth emphasizing here that this outcome is independent of the specific dynamics by which the flavor-mixing couplings $\lambda_{1,2}$ could have potentially originated near the scale $M$. Thus, the condition for resonance is satisfied self-consistently within the effective theory framework of the minimal CW model. The condition, however, worsens upon EWSB as tree-level corrections to the masses and mixings begin to dominate.

\begin{figure}[htb]
	\centering
\hspace{-1cm}
  \begin{subfigure}{0.2\textwidth}
%      \centering
	\includegraphics[scale=0.6,keepaspectratio=true]{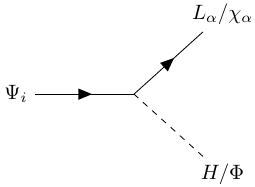}
%\hspace{3.5cm}         
      \caption{}	\label{fig:tree}
  \end{subfigure}%
  	%\hspace{-3.5cm}
  \begin{subfigure}{0.2\textwidth}
%      \centering
	\includegraphics[scale=0.6,keepaspectratio=true]{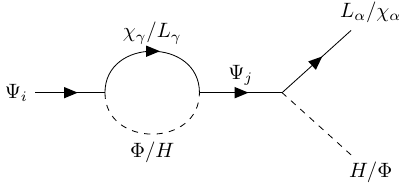}
      \caption{}	\label{fig:self}
  \end{subfigure}%
\caption{\textit{Decay channels of the heavy CW neutrinos: (a) Tree diagram and (b) Self-energy correction.}}\label{fig:decays}
\end{figure}

At temperatures above the EWSB scale the SM leptons and the dark fermions are massless and the heavy CW fermions can decay to a lepton and a Higgs doublet, or to a $\chi$ and $\Phi$ (Fig.\ref{fig:tree}). There are 16 physical phases in the model (see \ref{sec:appparamcount} for a discussion on the parameter counting). and the consequent CP asymmetry produced in the left-handed sector, emanating from the decays of the heavy neutral fermions and antifermions, is encapsulated by the parameter,
\beq 
\epsilon^{(L)}_{i \alpha} \equiv \frac{\Gamma\left( \Psi_{i} \to L_{\alpha} H\right)-\Gamma\left( \Bar{\Psi}_{i} \to \Bar{L}_{\alpha} H^{\dagger}\right)}{\sum_{f} \left[ \Gamma\left(\Psi_{i} \to f \right) + \Gamma\left(\Bar{\Psi}_{i} \to \Bar{f} \right)\right]},
\label{eq:aysymm1}
\eeq
where $f$ denotes all kinematically allowed final states and $\Psi_{\beta} \equiv \Psi^{L}_{\beta} + \Psi^{R}_{\beta}$. We call it the left asymmetry (denoted by the superscript $L$). Surely, an equal and opposite CP asymmetry arising from decays to the right-handed dark fermions is implied by the CPT theorem. Thanks to the localization of the SM leptons and the fields $\chi$ at two different boundaries of the CW lattice, the effective interaction between them is clockwork suppressed. Consequently, the right asymmetry does not equilibrate with the left asymmetry, enabling the latter to be independently utilised for discussing the leptogenesis mechanism.

The dominant contribution to $\epsilon$ would emerge from the interference of the tree-level and the self-energy corrected decay amplitudes (Fig.\ref{fig:decays}). The left asymmetry, therefore, can be expressed as,
\beq 
\begin{split}
\epsilon^{(L)}_{i \alpha}  \approx \frac{-1}{8\pi}&\sum_{j,\gamma}\left[\xi^{i \alpha}_{\gamma j}/\left(2|\Tilde{\lambda}_{1 \alpha i}|^2  + |\Tilde{\lambda}_{2 \alpha i}|^2 \right) \right] \\
& \times \frac{\left( m^2_{i} - m^2_{j}\right) m_i m_j}{\left[\left( m^2_{i} - m^2_{j}\right)^2 + \left( m_i \Gamma_i + m_j \Gamma_j\right)^2\right] } 
\end{split} \, .
\label{eq:asymm2}
\eeq
Here, $m_i$ denotes masses for the CW fermions $\Psi_i$. Also, $\Tilde{\lambda}^{1,2}_{\alpha i} \equiv \sum_{\beta}V^{L,R}_{\beta i}a^{L,R}_{(n=i, j=N,0)}\, \lambda^{1,2}_{\alpha \beta}$, where $V$ defines the unitary transformation from the flavor basis $(\beta)$ to the mass basis $(i)$ of the CW fermions, and the entity $\xi$ --- which acts as a Jarlskog invariant --- is defined as,
\beq \label{eq:asymm3}
\xi^{i \alpha}_{\gamma j} = \mbox{Im}\left[\left(2 \Tilde{\lambda}^*_{1 \alpha i} \Tilde{\lambda}_{2 \gamma i} \Tilde{\lambda}^*_{2 \gamma j} \Tilde{\lambda}_{1 \alpha j}\right)N \right].
\eeq
The factor $N$ accounts for the contribution of all decaying fermions in a particular generation of the CW spectrum. This is based on the reasonable approximation that their contributions are numerically similar as suggested by the distributions given in eq. \ref{eq:eigv}. The expression for the corresponding right asymmetry, generated by the decays $\Psi_i \to \chi_{R \alpha} \Phi$, is simply obtained by applying the interchange $\lambda_1 \leftrightarrow \lambda_2$ in eq.\ref{eq:asymm3} which implies $\epsilon^{(R)}=-\epsilon^{(L)}$, as expected. Clearly, the CP asymmetries $\epsilon^{(L,R)}$ vanish when either of the Yukawa matrices $\lambda_{1,2} \to 0$ or when $[\lambda_1, \lambda_2]=0$.

\subsection{Mechanism for leptogenesis}

At very high temperatures $(T \gtrsim m_{\Psi})$, the heavy CW fermions are, understandably, in thermal equilibrium with the SM sector for $\mathcal{O}(1)$ couplings. As previously noted, the decays of the heavy neutrinos to the left-handed leptons are responsible for the left leptonic asymmetry whereas those to the  right handed dark fermions generate an equal and oppositely signed right asymmetry. Therefore, the only processes which could potentially eliminate the net asymmetry are the inverse decays. Then, the foremost requirement for a lepton number preserving (Dirac) leptogenesis is to ensure that the asymmetry generating decays cease to equilibrate with the corresponding inverse decays before $B+L$ violating sphaleron transition rates begin to fall at the threshold temperature $T_{sp} \sim 150$ GeV~\cite{Rubakov:1996vz,KUZMIN198536}. In our case this implies that the heavy neutrinos must go out of equilibrium well above $T_{sp}$, following which all decays would contribute to the total lepton asymmetry. The left asymmetry thus generated cannot be further washed out as any of the active processes involving the leptons, dark fermions and the scalars would be fermion number conserving. Additionally, as argued before, all possible equilibration processes with the corresponding right asymmetry would be CW suppressed. The net lepton asymmetry in the left-handed sector would subsequently get converted to baryon asymmetry through sphaleron transitions. Typically, Dirac leptogenesis models contain suppressed couplings for the decay vertices due to their proportionality to the neutrino masses. This leads to a weak washout scenario where the heavy particles decouple from the thermal bath at temperatures not far from their masses, \emph{i.e.} at $z_D \equiv m/T_D  \sim \mathcal{O}(1)$, where $T_D$ is the decoupling temperature. For our model, however, the heavy fermion dynamics is practically decoupled from the neutrino mass generation which results in unsuppressed effective vertices for the decay processes. In this case the wash-out factor defined at the temperature $T \sim m_{\Psi}$ is,
\beq
\mathcal{K}_{i \alpha} \equiv \Gamma\left( \Psi_{i}\right)/\mathcal{H}(z=1) \ggg 1,
\eeq
where $z \equiv m_{\Psi} / T$ and $\mathcal{H}$ is the Hubble parameter. Such an extreme wash-out suggests late decoupling for the parent neutrinos. Therefore, to avoid spoiling the resonance condition they need to be heavy enough so as to reach the out-of-equilibrium state before EWPT occurs. This sets a lower bound $m_{\Psi} \gtrsim \mathcal{O}(10 \, \mbox{TeV})$ on the CW neutrinos. Since the decays to SM leptons (and $\chi$) are the dominant processes, roughly the entire $\Psi$ abundance is converted to lepton number once the washout rate is such that $\Gamma \lesssim \mathcal{H}$. 
In this case a conservative estimate of the total (left-handed) lepton asymmetry (produced by the $i$-th generation of the CW fermions once they decouple) is given by,
\beq
Y^{i}_{\Delta L} \approx \sum_{\alpha} \epsilon^{(L)}_{i \alpha} \mbox{Br}\left( \Psi_{i} \to L_{\alpha} H\right)Y^{i}_{\Psi, eq}(z^*).
\eeq
Here, $Y^{i}_{\Psi, eq}(z^*)$ is the equilibrium density of $\Psi^{i}$ at the epoch $z=z^*$ when it departs from its state of equilibrium with the SM leptons. 
Taking cue from \cite{Abada:2006ea}, the value of $z^*$ can be estimated by the relation
\beq 
z^*=\log{\sum_{\alpha}\mathcal{K}_{i \alpha}} + 5 \ln{(z^* /2)},
\eeq 
as the point of maximum contribution to $Y_{\Delta L}$ obtained by employing the steepest-descent method to evaluate the approximate integral equation,
\beq
Y^{i}_{\Delta L} \approx \sum_{\alpha} \epsilon_{i \alpha} \int_{0}^{\infty} dz \frac{K_1}{4g^*} z^2 e^{-\int_{z}^{\infty}dz'\left[z'^3/2\right]K_1 \mathcal{K}_{i \alpha}} \, .
\eeq 
Here, $K$ is a modified Bessel function of the second kind and $g^*$ denotes the effective degrees of freedom. For our scenario this engenders a solution $z^* \sim 20$. 

In order to determine what kind of textures of the Yukawa matrices could facilitate leptogenesis, we performed a scan using randomly generated values of the free parameters in the model (see \ref{sec:appparamcount} for details on the parameter counting). Referring to eq.\ref{eq:lam1} we see that the $\lambda_1$ matrix can be parametrised by the three active neutrino masses, the three mixing angles and one phase in the PMNS matrix $U_L$, as well as the three mixing angles and six phases in the unitary matrix $U_R$. $\lambda_2$, on the other hand is parameterised by nine real entries and nine phases. To comply with the neutrino oscillation data we fix the parameters of $U_L$ as per the global fit results of ref.\cite{Esteban:2020cvm}. Thus, in effect, we are left with a total of 12 real free parameters and 15 phases that parameterise the matrices $\lambda_{1,2}$. 
For the scan, in order to exhibit the role of the CW mechanism in reducing potential hierarchies in the matrix elements, we restrict to nearly $\mathcal{O}(1)$ entries for the matrices and set the real parameters to take random values distributed uniformly over the range $[0.01,1]$. For the phases we set the range $[0,2\pi]$. Then we fix the CW parameters $m$ and $q$ as well as the lightest active neutrino mass $m^{\nu}_0$, for which the appropriate value(s) of $N$ is deduced so as to be consistent with the neutrino masses and mixings. As a result of a scan over the Yukawa matrices for several benchmark values of $m$, $q$, $N$ and $m^{\nu}_0$, we found that a large fraction ($\sim 20 - 40 \, \%$) of the random textures generated pass the condition for a successful leptogenesis for a normal ordering of the neutrino masses\footnote{This is true for $q \lesssim 4$. We find that the fraction of the viable points decreases with an increase in the value of $q$ and falls below the $10\%$ mark for $q \gtrsim 5$. This can be understood from the fact that for a fixed set of hierarchies in the neutrino masses an increase in the value of $q$ requires a corresponding decrease in $N$. Since the eigenvectors in the CW sector induce a dependence on $N$ of the CP asymmetry ($\epsilon \propto N$) and of $z^{*}$, $|Y_{\Delta L}|$, too, effectively depends on $N$ and in fact decreases with decreasing $N$.}. As for the inverted ordering scenario, the scan shows a $\sim 10 - 30 \, \%$ success rate. Therefore, the observed neutrino masses and mixings are indeed compatible with the mechanism for leptogenesis for \emph{anarchic} textures of the Yukawa matrices having nearly $\mathcal{O}(1)$ entries --- with relatively more affinity for normal ordered neutrino masses. Interestingly, the success rates obtained in our analysis are comparable to those mentioned in ref.\cite{Hall:1999sn}, which focused on the possibility of anarchic mass matrices satisfying the prevalent neutrino data. This also goes to show that the results are robust with respect to the seemingly \emph{ad hoc} treatment of the flavor symmetry breaking in the model, \emph{i.e.} the leptogenesis mechanism is not sensitive to the details of the underlying theory as long as it is able to generate a flavor symmetric CW Lagrangian as well as localised flavor-mixing Yukawa couplings (one such possibility is described in \ref{sec:app5dcwm}).

For illustration, we summarise here the results of two benchmark cases for $m_{\Psi} \sim m q = 14$ TeV with different values of the CW parameters and $m_0^{\nu}$. Corresponding to each case we show in the following a specific sample of the $\lambda_1$ and $\lambda_2$ matrices with nearly $\mathcal{O}(1)$ entries, picked out of the many textures which satisfy the requirements for leptogenesis while being consistent with the neutrino oscillation data.

\underline{$q=2$, $N=42$:} 
Assuming normal ordering for the active neutrinos with the lightest neutrino mass $m^{\nu}_0 = 1.8 \times 10^{-5} \, \mbox{eV}$ we have the following viable textures for $\lambda_1$ and $\lambda_2$, respectively--- 
\[\left(
\begin{array}{ccc}
 0.160\, +0.075~ i & -0.060-0.013 ~i & -0.086-0.157 ~i \\
 0.281\, -0.485 ~i & 0.234\, -0.021 ~i & 0.133\, +0.756 ~i \\
 0.055\, -0.485 ~i & 0.291\, +0.076 ~i & 0.045\, +0.919~ i \\
\end{array}
\right)\]

\[\left(
\begin{array}{ccc}
 -0.344+0.572~ i & 0.150\, +0.411~ i & 0.751\, -0.400~ i \\
 0.640\, -0.143~ i & 0.700\, -0.027~ i & 0.073\, +0.549~ i \\
 0.644\, +0.904~ i & 0.562\, +0.341~ i & 0.613\, +0.687~ i \\
\end{array}
\right).\]
The flavor-wise CP asymmetries\footnote{Note that for each flavor, we quote here the asymmetry emanating from the entire tower of heavy neutrinos. The corresponding asymmetry per CW level $(n)$ is approximately given by $\epsilon_{i}/N$.} generated in this case are given by $\epsilon^{(L)}_i\equiv\sum_{\alpha}\epsilon^{(L)}_{i \alpha}= -\epsilon^{(R)}_i=\{-0.066, \, -1.012, 
  \, 0.0005\}$.
 
 Similar examples also exist for the inverted ordering scenario, e.g. with $m^{\nu}_0 = 6.5\times 10^{-5} \, \mbox{eV}$ we have for the $\lambda_{1,2}$ matrices---

\[
\left(
\begin{array}{ccc}
 -0.259+0.042~ i & -0.372+1.105 ~i & 0.108\, -0.774 ~i \\
 0.391\, -0.109~ i & 0.123\, -0.330~ i & 0.397\, -0.699~ i \\
 -0.473+0.066 ~i & -0.410+0.348 ~i & -0.274+0.792~ i \\
\end{array}
\right)
\]
\[\left(
\begin{array}{ccc}
 0.135\, -0.351 ~i & -0.496+0.449~ i & 0.577\, -0.929 ~i \\
 0.062\, -0.661~ i & 0.099\, -0.559~ i & -0.221-0.419~ i \\
 -0.889-0.387 ~i & 0.690\, -0.290~ i & -0.292-0.889 ~i \\
\end{array}
\right)\]
with $\epsilon^{(L)}_i = -\epsilon^{(R)}_i = \{ 0.371, \, 0.027,
    \, 0.487 \}$.

\underline{$q=4$, $N=21$:}
In this case, for normal ordering with $m^{\nu}_0 = 5.8 \times 10^{-5} \, \mbox{eV}$, we find the following $\lambda_{1,2}$ matrices to be viable--- 

\[
\left(
\begin{array}{ccc}
 -0.023-0.180 ~i & -0.081+0.022 ~i & 0.106\, -0.055~ i \\
 -0.295+0.818 ~i & 0.051\, +0.043~ i & 0.053\, -0.175~ i \\
 -0.352+0.860 ~i & 0.176\, +0.018~ i & -0.131-0.037~ i \\
\end{array}
\right)\]
\[\left(
\begin{array}{ccc}
 -0.589+0.224 ~i & -0.104+0.514 ~i & 0.401\, +0.893 ~i \\
 -0.304-0.100 ~i & 0.523\, +0.962 ~i & 0.756\, +0.795~ i \\
 0.455\, +0.517~ i & -0.343-0.889 ~i & 0.681\, -0.585 ~i \\
\end{array}
\right)\]
engendering $\epsilon^{(L)}_i=-\epsilon^{(R)}_i\{-0.575, \, -0.389, \, -0.635 \}$.

For inverted ordering with $m^{\nu}_0 = 2.4\times 10^{-5} \, \mbox{eV}$ we have for $\lambda_1$ and $\lambda_2$---

\[\left(
\begin{array}{ccc}
 0.027\, +0.046~ i & -0.763-0.439 ~i & -0.307+0.871 ~i \\
 -0.038-0.082 ~i & 0.539\, +0.302 ~i & -0.377+0.500 ~i \\
 0.030\, +0.095~ i & -0.483-0.495 ~i & 0.226\, -0.633~ i \\
\end{array}
\right)\]

\[\left(
\begin{array}{ccc}
 0.277\, +0.744~ i & -0.907-0.051~ i & 0.469\, +0.172 ~i \\
 -0.948-0.677 ~i & 0.924\, -0.426~ i & 0.875\, -0.183 ~i \\
 0.761\, -0.719~ i & 0.0744\, +0.066~ i & -0.382-0.974~ i \\
\end{array}
\right)\]
with $\epsilon^{(L)}_i=-\epsilon^{(R)}_i=\{-0.990, 
  \, 0.032, \, -0.420 \}$.

In all the above cases, a total lepton asymmetry $|Y_{\Delta L}| \sim \mathcal{O}(1) \times 10^{-10}$ is generated which lies in the right ballpark to generate the correct baryon asymmetry through sphaleron transitions \cite{Harvey:1990qw}.

\section{The Dark Sector} 

From the minimal setup for leptogenesis in the CW paradigm, we see the necessary existence of additional neutral fermions $\chi_{\alpha}$ coupling to the CW sector through the scalar $\Phi$. As previously mentioned, the lightest of these extra particles would be stable over the age of the Universe if $m_{\chi,\Phi} < m_{h}$, where $m_{h}$ is the SM Higgs mass. The case for $\chi$'s as DM is readily ruled out from overclosure as the dominant process for annihilation to the visible sector, $\chi_{i} \Phi \to \nu h$, mediated by $\Psi$'s is overly suppressed. E.g., for the range of $\lambda_{1,2}$ values consistent with leptogenesis and $m_{\Psi} \sim \mathcal{O}(10 \, \mbox{TeV})$, $\langle \sigma v \rangle_{max} \sim 10^{-31}~\mbox{cm}^3 \mbox{s}^{-1}$. The scalar $\Phi$, on the other hand, can saturate the relic abundance via $3 \to 2$ number changing processes. In this case, the SIMPy DM mass can be estimated to be $\mathcal{O}(10-100 \, \mbox{MeV})$ for $\lambda_{\Phi} \sim \mathcal{O}(1)$ \cite{Hochberg:2014kqa}. However, the secluded nature of the dark sector results in an extremely suppressed DM-SM interaction through the CW portal. Therefore, $\Phi$ as DM would struggle to be in kinetic equilibrium with the SM as it approaches freeze-out and, as a consequence, would heat up excessively. Such a scenario is severely constrained from the usual dynamics of structure formation in the Universe which suggests that the $\Phi-\chi$ system can only constitute a small fraction of the total DM abundance. This is easily achieved in the minimal model for $\lambda_{\chi,\Phi} \sim \mathcal{O}(1)$, $ m_{\Phi} \lesssim m_{\chi} < \mathcal{O}(1 \, \mbox{MeV})$ which enhances the rates for $\chi \chi \to \Phi \Phi$ and $3\Phi \to 2\Phi$ by the factors $\frac{\langle \sigma v\rangle}{\langle \sigma v\rangle_0}\Big|_{2 \to 2} \gtrsim 10^2$ and $\frac{\langle \sigma v^2\rangle}{\langle \sigma v^2\rangle_0}\Big|_{3 \to 2} \gtrsim 10^5$, respectively, where $\langle \sigma v\rangle_0$ and $\langle \sigma v^2\rangle_0$ denote typical threshold values for the thermally averaged annihilation cross-sections required to obtain the correct relic abundance. Clearly, this would deplete the $\chi-\Phi$ abundances to a negligible percentage of the required DM yield. With the dynamics of leptogenesis being virtually independent of the details of the dark sector (except for the masses), it might be possible to juxtapose the basic construction proposed here with an extended dark sector, ostensibly with a heavy DM candidate of mass $\sim \mathcal{O}(m_{\Psi})$ so as to have enhanced DM-SM interactions. Such explorations, however, are beyond the purview of this work.

\section{Discussion and conclusion}
We have presented a minimal model within the clockwork paradigm where the correct lepton asymmetry for baryogenesis is naturally produced, while accounting for small Dirac masses for the active neutrinos, with nearly $\mathcal{O}(1)$ and anarchic Yukawa couplings. It has a few distinct, yet interesting, features. For one, the suppression in the light neutrino masses as well as in the equilibration between the CP asymmetries of the left- and the right-handed sectors is facilitated by the localization of the visible and the dark sectors on the CW lattice and, therefore, is practically independent of the heavy mass scale $m_{\Psi}$. Secondly, the CW mechanism stipulates that couplings of the heavy states to the leptons are roughly of the order of the Yukawa matrix elements, in stark contrast with the seesaw based \textit{neutrinogenesis} scenarios where the effective couplings are proportional to the neutrino masses. Therefore, for Dirac leptogenesis, where self-energy corrections are the dominant higher order contributions, $\mathcal{O}(1)$ Yukawa couplings warrant a resonant enhancement in the CP asymmetry to counter the characteristically strong washout effects. Of further assistance are the contributions due to all the heavy neutrinos present in the CW spectrum, which tend to enhance the net CP asymmetry by roughly an order of magnitude for $N \sim \mathcal{O}(10)$.

The necessary small mass-splittings for resonance are naturally achieved through loop-induced flavor mixings caused by the explicit breaking of the flavor $\mathbb{Z}_3$ symmetry by random textures of the Yukawa matrices. Our study shows that the leptogenesis mechanism must operate well above the EWSB scale, which is when the resonance condition is satisfied. This engenders a lower limit on the heavy fermion masses, namely $m_{\Psi}\gtrsim \mathcal{O}(10 ~\rm TeV)$.  An immediate consequence of this limit is that the model trivially evades constraints from charged LFV processes $\ell_i\to \ell_j\gamma$, $\ell_i\to \ell_j \ell_k \ell_l$. On a related note, the CW scale in the model is clearly beyond the reach of the LHC as well as some of its upcoming derivatives. However, it could be probed in the future at some of the proposed energy frontier experiments, e.g. at the HE-LHC, FCC and the multi-TeV muon collider. Note that even if a SUSY embedding of the model is attempted, the lower limit obtained here on the scale of leptogenesis would be consistent with the upper limit on the reheating temperature occasioned by gravitino abundance, viz. $T_{reh}< 10^7- 10^9$ GeV~\cite{Buchmuller:2004tu, KHLOPOV1984265, ELLIS1984181}.

In conclusion, we would like to remark that the generation of hierarchically suppressed couplings or mass scales is a characteristic feature of the clockwork mechanism which has led to its application in phenomenology largely in that context alone. Importantly, in most models based on the CW mechanism the dynamical heavy degrees of freedom (CW \textit{gears}) had but only a passive role in the phenomenology discussed therein \cite{Alonso:2018bcg,Kehagias:2016kzt}. In this work, we have shown a more active, albeit pivotal, role for the CW gears in the context of leptogenesis. This not only paves the way for further explorations related to BAU within the CW framework, but also in other theory space constructions based on a localization mechanism, see e.g. ref.\cite{Craig:2017ppp}.

\section*{Acknowledgements}
We are grateful to Ananya Mukherjee and Mathew Thomas Arun for drawing our attention to this problem. We also thank Debajyoti Choudhury for the illuminating discussions and his feedback on the manuscript, and Rick S. Gupta and Tuhin S. Roy for a useful comment on an earlier version of the manuscript.
S.M. acknowledges research Grant No. CRG/2018/004889 of the SERB, India. T.S. would like to acknowledge the support from the Dr. D.S. Kothari Postdoctoral fellowship scheme no. F.4-2/2006 (BSR)/PH/20-21/0163.

\appendix

\section{Parameter counting in the Yukawa sector} \label{sec:appparamcount}
The number of independent physical parameters in the Yukawa sector is most succinctly determined by looking at the pattern of the full flavor symmetry breaking in the theory when Yukawa couplings are present. In the absence of the Yukawa couplings, the fermion sector (including the SM charged leptons as well) in our model has the flavor symmetry $U(3)^3 \times U(3N+3)^2$. Now, introducing the most general Yukawa couplings (\emph{i.e.} without imposing a $\mathbb{Z}_3$ symmetry in the CW sector), while complying with the charge assignments of the fields, breaks the flavor symmetry to an overall lepton number symmetry $U(1)_{lep}$. The corresponding real and phase broken generators are given by,
\beq
B_{real}=9N^2+15N+15 ,\, \, B_{phase}=9N^2+21N+29.
\eeq
On the other hand, the number of real and phase parameters in the Yukawa couplings amount to,
\beq
Y_{real}=Y_{phase}=9(N+2)^2.
\eeq
Therefore, the total number of real and phase physical parameters in the general flavor-mixing Lagrangian for our scenario is
\beq
\begin{split}
&{\rm Real}=Y_{real}-B_{real}=21(N+1)\\
&{\rm Phase}=Y_{phase}-B_{phase}=15N+7.
\end{split}
\eeq
Understandably, for our model realisation with a $\mathbb{Z}_3$ flavor symmetry in the CW sector and the localized flavor-mixing Yukawa couplings we have fixed the values of a large subset of the total parameter count. Thus, in order to focus on the relevant phenomenology, we have limited the number of free physical parameters in the model to the following--- the CW parameters $m$ and $q$; the matrix $\lambda_1$ parameterised by the 3 light neutrino masses, 6 mixing angles and 7 phases; the matrix $\lambda_2$ parameterised by 9 real entries and 9 phases; and the matrix $\lambda_{\chi}$ with diagonal entries determined from the masses of the three $\chi$ fermions.

\section{A 5D clockwork model} \label{sec:app5dcwm}
The clockwork model being discussed can be deemed as the deconstruction limit of a five-dimensional theory with bulk fermions in a linear dilaton (LD) geometry defined by the metric \cite{Antoniadis:2011qw},
\beq
\begin{split}
ds^2&=g_{MN}dx^{M}dx^{N}\\
&=e^{2 \sigma(z)}\left( \eta_{\mu \nu}dx^{\mu}dx^{\nu} + dz^2\right) \, .
\end{split}
\eeq
Here, $x^{\mu}$ and $\eta_{\mu \nu}$ denote the flat 4D coordinates and metric, respectively. $z$ stands for the fifth coordinate which is compactified on a $S^1/\mathbb{Z}_2$ orbifold of size $\pi R$ with a 3-brane in each of its boundaries. The warp profile $\sigma$ is given by,
\beq
\sigma(z)=-\frac{2}{3}k|z|,
\eeq
where, $k$ is a bulk mass parameter. For illustration, a 5D realisation of the effective CW theory can be imagined as follows. In the LD background, the flavor-symmetric 5D bulk action for three generations of a fermion can be expressed as,
\beq
\begin{split}
    \mathcal{S}_{Bulk}=\sum_{\beta=1}^{3}\int d^{4}x \, dz \sqrt {-g}\big\{ -&\frac {i} {2}E_{a}^{\, M}\bar {\psi^{\beta}}\Gamma^{a}\overleftrightarrow{\partial_{M}}\psi^{\beta}  \\
    & - e^{-\sigma(z)}m_{\psi}\bar{\psi}^{\beta}\psi^{\beta} \big\},
\end{split}
\eeq
with a mass parameter $m_{\psi}$ and
\beq
\Gamma^{a}\equiv\left\{ \gamma ^{\mu },i\gamma ^{5}\right\} , \quad \overleftrightarrow {\partial _{M}}\equiv\overrightarrow {\partial}_M-\overleftarrow {\partial }_M \, . 
\eeq
$E^{M}_{a}=e^{-\sigma(z)}\delta^M_a$ specifies the requisite vielbien for this case. As prescribed in ref.\cite{Giudice:2016yja}, a deconstruction of the bulk action --- by discretising the fifth dimension into a lattice of $N+1$ sites with a characteristic spacing $a=\pi R / N$ --- would lead to the CW theory in eq.\ref{eq:lagcw}. One could also introduce interaction terms on the branes where the visible and dark sectors are localised, namely,
\beq \label{eq:branevis}
\begin{split} 
\mathcal{S}_{vis.}= \int & d^4x \, dz \sqrt{\frac{-g}{g_{zz}}} \Big[ \sum_{\alpha=1}^{3}\mathcal{L}^{\alpha}_{vis.} \\
&- \sum_{\alpha,  \beta}c^{v}_{\alpha, \beta}\Bar{L}_{\alpha}\Tilde{H}\psi^{\beta}(z)\Big]\delta(z-\pi R)
\end{split}
\eeq
\beq \label{eq:braneds}
\begin{split}
\mathcal{S}_{d.s.}= \int  d^4x \, dz &\sqrt{\frac{-g}{g_{zz}}} \Big[ \sum_{\alpha=1}^{3}\mathcal{L}^{\alpha}_{d.s.} \\
&- \sum_{\alpha,  \beta}c^{d}_{\alpha, \beta} \Phi \Bar{\psi}^{\beta}(z)\chi_{R \alpha}\Big]\delta(z) \, .
\end{split}
\eeq
In the preceding, $c^{v,d}$ are effective couplings which break the flavor symmetry explicitly and map onto the $\lambda_{1,2}$ couplings in eq.\ref{eq:lagcwsm} upon discretisation. The origin of $c^{v,d}$ can be attributed to the VEVs acquired by 9 heavy complex scalar fields (\emph{flavons}\footnote{Not to be confused with the flavon field usually encountered in Froggatt-Nielsen-type scenarios.}) $\phi^{(\alpha, \beta)}$ each in the visible and dark sector branes. To see this systematically, let us endow each bulk fermion family with a charge under a global flavor symmetry $U(1)_{\beta}$, and each SM lepton and $\chi$ family with a charge under $U(1)^{(l)}_{\alpha}$ and $U(1)^{(\chi)}_{\alpha}$, respectively. Furthermore, the brane-localised flavons can be assigned charges under $U(1)_{\alpha} \times U(1)_{\beta}$ such that, before SSB in the $\phi$ sector, we have dimension-five terms on the branes of the form
\beq
\begin{split}
\mathcal{S}_{vis.}\supset \int & d^4x \, dz \sqrt{\frac{-g}{g_{zz}}}\\
\times &\Bigg[- \sum_{\alpha,  \beta}\frac{\phi_{vis.}^{(\alpha, \beta)}}{\Lambda}\Bar{L}_{\alpha}\Tilde{H}\psi^{\beta}(z)\Bigg]\delta(z-\pi R)
\end{split}
\eeq
and
\beq
\begin{split}
\mathcal{S}_{d.s.}= \int  d^4x \, dz &\sqrt{\frac{-g}{g_{zz}}} \\
\times & \Big[
- \sum_{\alpha,  \beta}\frac{\phi_{d.s.}^{(\alpha, \beta)}}{\Lambda} \Phi \Bar{\psi}^{\beta}(z)\chi_{R \alpha}\Big]\delta(z) \, ,
\end{split}
\eeq
where $\Lambda$ denotes a cutoff for the theory. Now, a configuration of the potential for the flavons can be arranged wherein they acquire VEVs such that  $\langle \phi_{vis.}^{(\alpha, \beta)} \rangle/\Lambda = c^{v}_{\alpha, \beta}$ and  $\langle \phi_{d.s.}^{(\alpha, \beta)} \rangle/\Lambda = c^{d}_{\alpha, \beta}$, thereby spontaneously breaking the flavor symmetry. Upon integrating out the heavy flavons\footnote{Note that the resulting Goldstones can be made appropriately massive by introducing explicit $U(1)_{\alpha}\times U(1)_{\beta}$ breaking terms in the brane actions.} we finally obtain the actions of eqs.\ref{eq:branevis} and \ref{eq:braneds}. Note that the characteristic scale of the flavor SSB in the 5D model would manifest as a cutoff akin to the scale $M$ mentioned in the discrete model (eq.\ref{eq:splitloop}).

\bibliographystyle{apsrev4-1} 
\bibliography{v1.bib}

\end{document}